\documentclass[12pt]{iopart}
\usepackage{units}
\usepackage{amssymb}
\usepackage{graphicx}

\makeatletter

\providecommand{\tabularnewline}{\\}

\usepackage{iopams}
\usepackage{setstack}
\makeatother

\begin{document}

\title{Magnetoresistance of composites based on graphitic disks and cones}

\author{Jozef \v{C}ern\'{a}k$^1$, Geir Helgesen$^2$, Fredrik Sydow Hage$^2$\footnote{Present Address: SuperSTEM Laboratory, SciTech Daresbury, Daresbury,
WA4 4AD, United Kingdom}, Jozef Kov\'{a}\v{c}$^3$}

\address{$^1$P. J. \v{S}af\'{a}rik University in Ko\v{s}ice, Institute of Physics,
Jesenn\'{a} 5, SK-04000 Ko\v{s}ice, Slovak Republic} \ead{jozef.cernak@upjs.sk}

\address{$^2$Institute for Energy Technology, Physics Department, NO-2007 Kjeller,
Norway and Department of Physics, University of Oslo, NO-0316 Oslo,
Norway}

\address{$^3$Institute of Experimental Physics, Slovak Academy of Sciences, Laboratory
of nanomaterials and applied magnetisms, Watsonova 47, SK-04000 Ko\v{s}ice,
Slovak Republic}

\begin{abstract}
We have studied the magnetotransport of conical and disk-shaped nanocarbon
particles in magnetic fields $\left|B\right|\leq9\:\mathrm{T}$ at
temperatures $2\leq T\leq300\:\mathrm{K}$ to characterize electron
scattering in a three dimensional disordered material of multilayered quasi
2D and 3D carbon nanoparticles.  The microstructure of the particles was
modified by graphitization at temperatures $1600^{\circ}\mathrm{C}$ and
$2700^{\circ}\mathrm{C}$.  We find clear correlations between the
microstructure as seen in transmission electron microscopy and the
magnetotransport properties of the particles.  The magnetoresistance
measurements showed a metallic nature of samples and
positive magnetoconductance which is a signature of weak localization in disordered
systems.  We find that the magnetoconductance  at
low temperatures resembles quantum transport in single-layer graphene
despite the fact that the samples are macroscopic and three dimensional,
consisting of stacked and layered particles, which are
randomly oriented in the bulk sample.  This graphene-like behaviour is
attributed to the  very weak interlayer coupling between the
graphene layers.

\end{abstract}
\submitto{\JPD}
\maketitle

\section{Introduction}

The electronic properties of the carbon allotropes, such as nanotubes,
graphene \cite{Castro,Peres,Sarma} and graphite \cite{Soule1958} are controlled by the object dimensionality \cite{Lee1985},
microscopic structure, disorder \cite{Lee1985,Imry,Altshuler1980},
charge carrier type, density of carriers
and their mobility \cite{Castro,Peres,Sarma}, temperature and external electric or magnetic
fields \cite{Castro,Peres,Sarma,Vora,Menon,Vavro}.

Electronic transport in conventional 2D systems structures
or thin films with magnetic impurities was explained by Hikami \textit{et
al.}\cite{Hikami}. For 3D systems this theory  was extended by Kawabata
\cite{Kawabata}.

Recent studies of graphene-like materials such as bi-layer graphene
and modified multi-layer graphene \cite{Tikhonenko2008,Baker2012,SevakSingh2012,Meng2012}
have shown both weak localization and weak antilocalization phenomena
depending on the sample preparation. The differences may partly be
attributed to variations in the stacking of the graphene layers from
the normal AB or Bernal type toward twisted commensurate layers which
was reported \cite{Meng2012} to enhance interlayers couplings and
scattering. Theoretical modeling \cite{Sarrazin} has shown that twisting
of bilayer graphene, e.g., by twist angle $\theta=21.8^{\circ}$, can
give commensurate structures and strongly modified interlayer coupling.
Producing bilayer or multilayer graphene with controlled twist angles
is an extremely challenging task. However, there may exist naturally
occurring multilayered graphene or pyrolytic graphite-like materials
that can resemble such configurations. Possible candidates that may
contain twisted graphene are the carbon cone particles \cite{Krishnan}.
These are conical or disk-shaped graphitic-like particles, and it
has been reported that certain of these show edge faceting which might
be consistent with an alternating twist angle of about $22^{\circ}$
between adjacent layers \cite{Garberg}. Our motivation for the current
study was to see how this conical topology influences the electronic
scattering mechanisms as may be recorded in magnetoresistivity. We
investigate resistance versus temperature and conductance versus magnetic
field of a powder of nanoparticles that was bound into mm-sized samples
using polymer binder. The nanocarbon particles were mainly cone- and
disk-shaped and were prepared with varying degree of graphitization.
These nanocarbon-polymer composite samples are apparently similar to granular
conductors \cite{Fung, Kuznetsov}.

The conductance versus magnetic field of the heat treated (HT) samples
displayed features of 2D transport. We find that the magnetoconductivity model
developed in the theory of McCann \textit{et
al.} \cite{McCann} for 2D graphene is suitable for discussing
our experimental data at low temperatures \cite{Wu2007}. We can
partially explain the observed behaviour as effect of very weak coupling
between misoriented layers \cite{SevakSingh2012,LopesdosSantos2007,Latil2007} inside
the disks and cones.

\section{\label{sec:Sample_description}Sample descriptions}

\subsection{\label{sub:Carbon-nanopowder}Nanocarbon particles}

The graphitic-like carbon powder was produced by the so called ``Kvaerner
Carbon Black and Hydrogen Process'' \cite{Hugdahl} which is an industrial,
pyrolytic process that decomposes hydrocarbons into hydrogen and carbon
using a plasma torch at temperatures above $2000^{\circ}\:\mathrm{C}$.  The as produced,
"raw" powder (in the following denoted HT-0) consists
of flat nanocarbon disks, open-ended carbon cones, and a
small amount of carbon black-like structures as seen in electron microscopy images
\cite{Krishnan,Garberg,Svasand,Heiberg-Andersen_2011}.  In order to improve the
crystalline quality of the particles, additional heat treatment was done at
high temperatures in an argon atmosphere for three hours, followed by a slow
natural cooling.  In this study, heat treatment was done at either
$1600^{\circ}\:\mathrm{C}$ (HT-1600) and $2700^{\circ}\:\mathrm{C}$ (HT-2700).

\begin{figure}
\includegraphics[width=7.5cm]{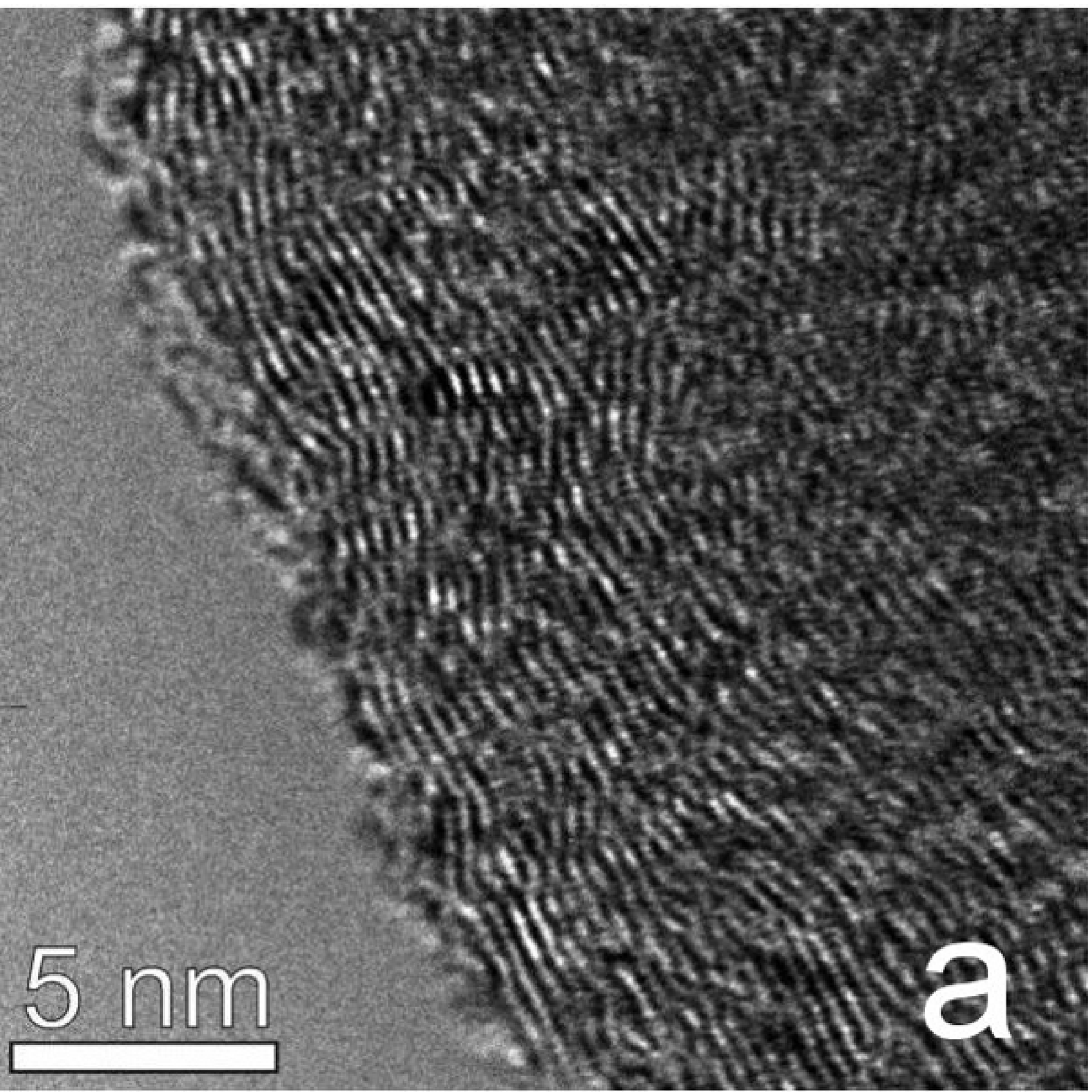}
\includegraphics[width=7.5cm]{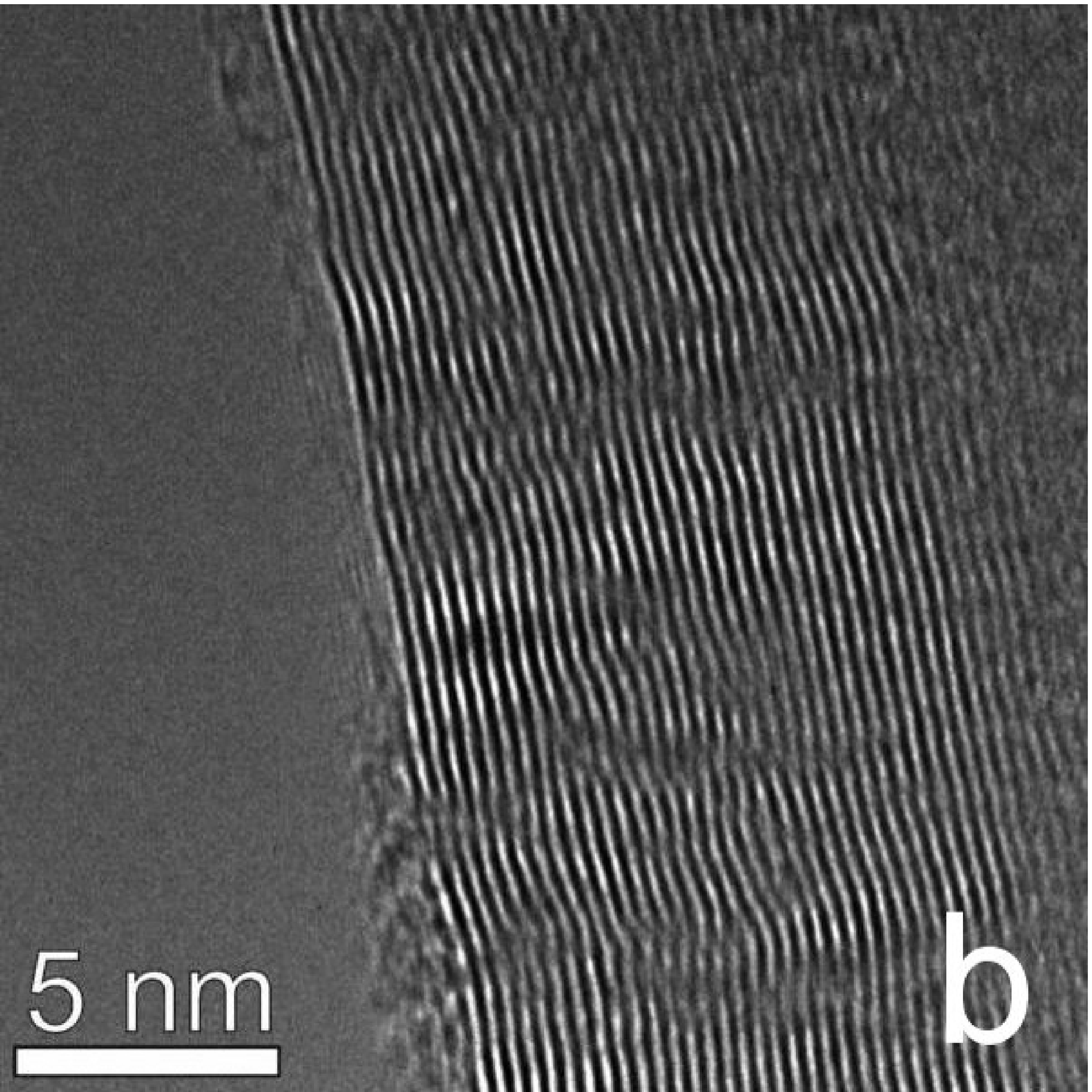}
\caption{\label{fig:TEM} High-resolution
transmission electron micrographs showing the microstructure of a) carbon
cones raw material and material heat treated at b) $1600^{\circ}\mathrm{C}$,
and c) $2700^{\circ}\mathrm{C}$.} 
\end{figure}

The carbon disks and cones exhibit a wide range of
diameters ($500-4000\:\mathrm{nm}$) with wall thicknesses of typically
$10-30\:\mathrm{nm}$.  However, particles with thickness in the range
$5-70\:\mathrm{nm}$ can be found.  Krishnan\textit{ et al}.  \cite{Krishnan}
showed that the observed macroscopic cone apex angles correspond to those of
perfect graphene cones with apex angles $\alpha=112.9^{\circ}$,
$83.6^{\circ}$, $60.0^{\circ}$, $38.9^{\circ}$, and $19.2^{\circ}$.  Perfect
graphene cones are defined by $n=1-5$ pentagonal disclinations incorporated
in close proximity in a graphene sheet.  Upon extension, the flat disks can
be considered as cones with $n=0$, i.e., with no pentagons.  Some of the
disks and the cones with a $112.9^{\circ}$ apex-angle showed six-fold and
five-fold faceting, respectively \cite{Naess}.  Transmission electron
microscopy (TEM) selected area diffraction (SAD) patterns of a disc and cones in
\cite{Garberg, Naess} exhibit concentric continuous rings including as set of distinct
spots with six-fold rotational symmetry and were interpreted in terms of
discs comprising a highly crystalline graphitic core enveloped by much
thicker layer of disordered carbon. The thickness of the crystalline core
in the as-produced disks was estimated to be only $10-30\%$ of the total
disk thickness \cite{Naess}.  Recent high-resolution TEM (HR-TEM)
micrographs \cite{Hage-1} clearly show that the outer enveloping layer of
the as-produced disks and cones is turbostratic.  Upon heat treatment, the
graphitic order increases with increased heat treatment temperature.  From
HR-TEM images \cite{Hage-1} and SAD patterns \cite{Garberg,Hage-1} it is
clear that at $2700^{\circ}\:\mathrm{C}$, the structure of the cone and disk
envelope can be described by partially overlapping extended graphitic
domains wherein graphene layers lack a well defined stacking order.  In the
present work, HR-TEM micrographs were acquired with a Jeol JEM 2010F
microscope operated at 200 kV.  Figure
\ref{fig:TEM} shows details of the cone structure at the edge
cones heat treated at $1600^{\circ}\mathrm{C}$ and
$2700^{\circ}\mathrm{C}$.  After heat treatment at
$1600^{\circ}\:\mathrm{C}$ the enveloping carbon layer remains turbostratic.
However, at $2700^{\circ}\:\mathrm{C}$ extended
graphitic domains are clearly present, which indicates a polycrystalline
rather than a purely turbostratic structure.  While the cone envelope shows
varying degrees of disorder depending on HT, the overall cone morphology is
dictated by the highly crystalline core.  In light of the reported
conformity of measured cone apex angles to those expected for idealised
graphene cones, it follows that the crystalline cores must conform to a
$n=1-5$ disclination topology.  Thus the entire structure of these particles
are to a first approximation referred to as carbon cones, whilst acknowledging
that at the nanoscale the cone structure differs somewhat from that expected
for a perfect multilayer graphene cone, depending on heat treatment
temperature.  Note also that these cones are intrinsically different from
such structures as the conical graphite crystals reported by Gogotsi
\textit{et al.} \cite{Gogotsi-2002} and carbon nanohorns \cite{Iijima-1999}.

The investigated composite samples show disorder on at least two length scales. On
the nanometer scale they are a mixture of crystalline parts, likely
containing many dislocations, grain boundaries and other defects, and
non-crytalline/amorphous matter.  On the micrometer scale the grainy nature
of the powder will make different packings of particles and form a solid
sample with a locally varying material density.

The crystallinity of the nanocarbon powder was investigated by powder x-ray
diffraction using a Bruker AXS D8 Advance diffractometer with
$\mathrm{CuK}_{\alpha}$ ($\lambda=1.5418\:\mathrm{\mathring{A}}$) radiation
and a LynxEye detector.  The powder was filled in $1.5$ mm glass
capillaries.  The interlayer distance was found to be
$\frac{c}{2}=3.53\pm0.03 \:\mathrm{\mathring{A}}$, $3.464\pm0.015 \:\mathrm{\mathring{A}}$, and
$3.373\pm0.001\:\mathrm{\mathring{A}}$ for the HT-0, HT-1600, and HT-2700 samples,
respectively.  The typical size of the crystalline domains was estimated
from the full-width at half-maximum of the diffraction peaks using the
Scherrer equation \cite{Warren}.  The coherence
length perpendicular to the graphene layers, $L_c$, increased from the range $1-3$
nm for the raw material (HT-0) to $4.5\pm0.8$ nm and $38\pm6$ nm for the samples annealed
at $1600^{\circ}\mathrm{C}$ and $2700^{\circ}\mathrm{C}$, respectively.
This means that in particular the heating to
$2700^{\circ}\mathrm{C}$ had a significant influence on the crystalline
order, in qualitative agreement with the HR-TEM images in figure
\ref{fig:TEM}.  Naess \textit{et al.} \cite{Naess} reported that the
in-plane coherence length $L_a\approx 20$ nm for samples similar to our
HT-2700 material.  These authors also reported a faceting along the edges of
the disks and the 5-pentagon cones into pairs of sectors of about
22$^{\circ}$ and (60-22)$^{\circ}$= 38$^{\circ}$, possibly reflecting an
underlaying twisted hexagonal structure.  Earlier it has been shown that an
alternating shift of the graphene in-plane axis of 21.8$^{\circ}$ between
subsequent layers in a conical or helical cone structure may give an optimal
graphitic alignment between the layers \cite{Eksioglu,Amelinckx}.  It was
proposed \cite{Naess} that similar alternating layer rotations may exist in
the particles of the current carbon material.

In a previous study of magnetic properties of raw nanocarbon powder
\cite{Cernak2013}, we have determined a residual amount of Fe using the
particle-induced x-ray emission method.  In the  HT-1600 powder,
the content of Fe is $<75\:\mu\mathrm{g/g}$.  In HT-2700 powder, the content
of Fe is lower than the method can detect ($<3\:\mu\mathrm{g/g}$)
\cite{Cernak2013}.

\subsection{\label{sub:Polymerized-carbon-powder}Bulk sample preparation }

In order to prepare a dense bulk sample, carbon powder was mixed with
polymethyl methacrylate (PMMA) dissolved in chloroform where the volume
of nanocarbon filler was $>60$ vol-\%. The resulting viscous solution
was placed on a mica substrate of thickness $\sim30\:\mathrm{\mu m}$
having a four wire arrangement for resistivity measurements, figure
\ref{fig:RvsT} (a), inset. The polymerization took place at room temperature.
Typical size of the polymerized sample was about $6\times6\times0.5\:\mathrm{mm}$
(l $\times$w $\times$t), i.e., the samples were macroscopic three
dimensional objects. The reproducibility of the sample preparation
method was verified by resistance measurements at room temperature
for the same four wire arrangement (figure \ref{fig:RvsT} (a), inset)
and other samples with the same sample dimensions. Typical sample
resistance was in the range $R=1-15\:\Omega$. The volume fraction
of carbon powder in the sample was much higher than the percolation
threshold \cite{Stankovich}, i.e., the sample conductance did not
depend qualitatively on the fraction of nanoparticles. Thus, the samples
were 3D bodies of randomly stacked disks (quasi 2D objects) and
cones (3D objects) with a structure similar to granular conductors
\cite{Fung, Kuznetsov}.

\section{\label{sec:Resistance-measurement}Resistance measurement}
The polymer-bonded nanocarbon samples were placed in the vacuum chamber
of a Quantum Design Physical Property Measurement System (PPMS). The
standard four wire method shown in the inset of figure \ref{fig:RvsT}
(a) was used to measure the resistance. The samples were driven by
an\textit{ ac} current of $10\:\mu\mathrm{A}$ in a resistivity measurement
mode of the PPMS. Resistance measurements were performed at several
temperatures $2\leq T<300\:\mathrm{K}$ in magnetic fields $-9\leq B\leq9\:\mathrm{T}$
applied perpendicular to the sample surface. We have verified that samples show linear
volt-ampere dependence at $T=300\:\mathrm{K}$ for currents  $ 1\leq I\leq100\:\mu\mathrm{A}$.

\section{\label{sec:Experimental-results}Experimental results}

The experimental results are presented using either the resistance
$R$ (section \ref{sub:Resistance-R-versus-T}) or the conductance $G$
(section \ref{sub:Conductance-G-versus-T}) of the nanocarbon-polymer
samples. Resistivity $\rho$ of the samples was determined by taking
into account their physical dimensions.

\subsection{\label{sub:Resistance-R-versus-T}Resistance versus temperature measurements}

\begin{figure}
\includegraphics[width=6.5cm]{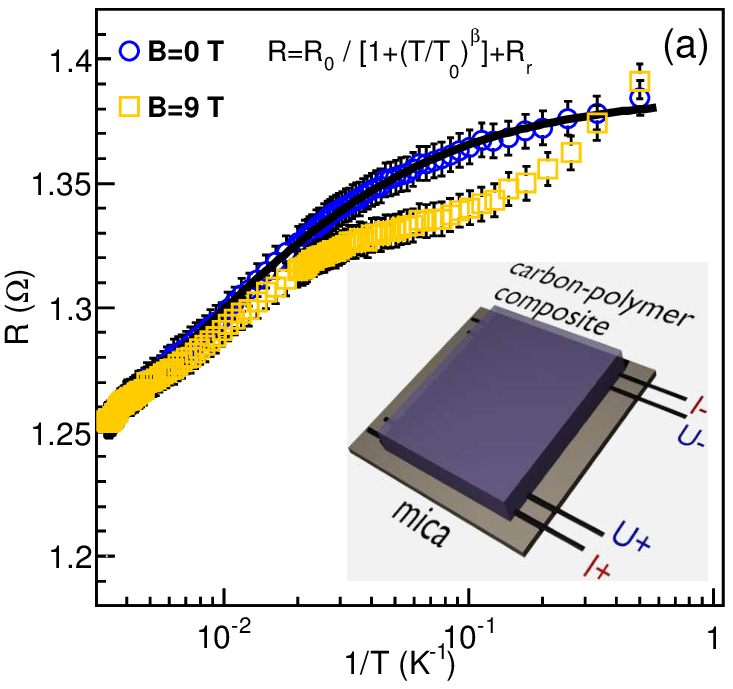}
\includegraphics[width=6.5cm]{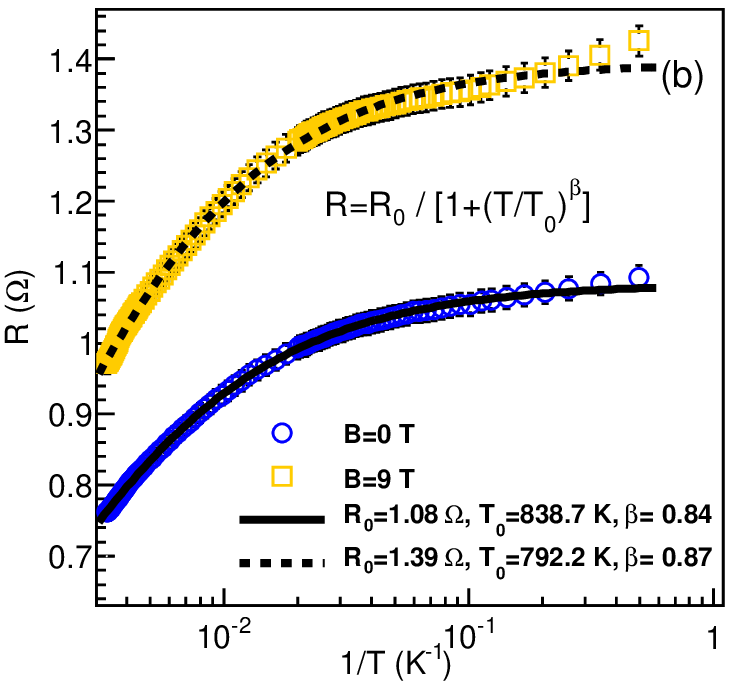}
\caption{\label{fig:RvsT}The resistance $R$ versus temperature
$T$ for magnetic fields $B=0$ and $9\:\mathrm{T}$.  (a)
(log-lin plot) polymer sample containing nanocarbon powder heated
at $1600^{\circ}\mathrm{C}$ (HT-1600) and (b) (log-lin plot) polymer
sample containing powder heated at $2700^{\circ}\mathrm{C}$ (HT-2700).
The inset shows the sample design and four-wire geometry used to measure
resistance. The fits to the experimental data without magnetic field
are shown as solid lines. In (b), the case with applied field $B=9\:\mathrm{T}$,
the fit is shown using a dashed line.}
\end{figure}

The temperature dependence of the resistance $R(T)$ of the HT-1600
sample without a magnetic field is shown in figure
\ref{fig:RvsT} (a). In the broad temperature interval $2\leq T<300\:\mathrm{K}$
the experimental data in figure \ref{fig:RvsT} (a) are approximated by:

\begin{equation}
R(T)=\frac{R_{0}}{1+\left(\frac{T}{T_{0}}\right)^{\beta}}+R_{r}\label{eq:res_1600}
\end{equation}
where $R_{0}$, $R_{r}$, $T_{0}$, and $\beta$ are parameters \cite{Vavro}.
These parameters are presented in table \ref{tab:Rfit}. The residual resistance $R_{r}$
had to be included due to a relatively high sample resistance at room temperature \cite{Vavro}.

In magnetic field $B=9\:\mathrm{T}$, the temperature dependence of
the resistance $R(T)$ shows a weak negative MR in the range from
room temperature $T\approx300\:\mathrm{K}$ down to $T\approx50\:\mathrm{K}$.
For lower temperature $10<T<50\:\mathrm{K}$, the negative MR due
to the magnetic field is more evident. However at temperature $T\approx3\mathrm{K}$
the MR changes sign and at temperatures $T<3\:\mathrm{K}$ it is positive.

The temperature dependence of the resistance $R(T)$ of the HT-2700
sample in figure \ref{fig:RvsT}(b) may also be approximated by equation
(\ref{eq:res_1600}) but with the residual resistance $R_{r}=0$. The
equation approximates experimental data both without and with magnetic field
in the broad temperature interval $2\leq T<300\:\mathrm{K}$
(see table \ref{tab:Rfit}). In magnetic field $B=9\:\mathrm{T}$,
a small deviation from equation (\ref{eq:res_1600}) was observed for low
temperatures $2\leq T<10\:\mathrm{K}$.

\begin{table}
\caption{\label{tab:Rfit}Parameters extracted in fits of the carbon nanoparticle
resistances to equation (\ref{eq:res_1600})
(HT-1600 and HT-2700 samples).}
\begin{tabular}{cccccc}
\hline
Sample & $B$ $(T)$ &  $\beta$ & $R_{0}\:(\Omega)$ & $T_{0}\:(K)$ & $R_{r}\:(\Omega)$\tabularnewline
\hline
HT-1600 & 0 & 0.84 & 0.19 & 130.7 & 1.19\tabularnewline
HT-2700 & 0 & 0.84 & 1.08 & 838.7 & 0\tabularnewline
HT-2700 & 9 & 0.87 & 1.39 & 792.2 & 0\tabularnewline
\hline

\end{tabular}
\end{table}

\begin{figure}
\includegraphics[width=7.5cm]{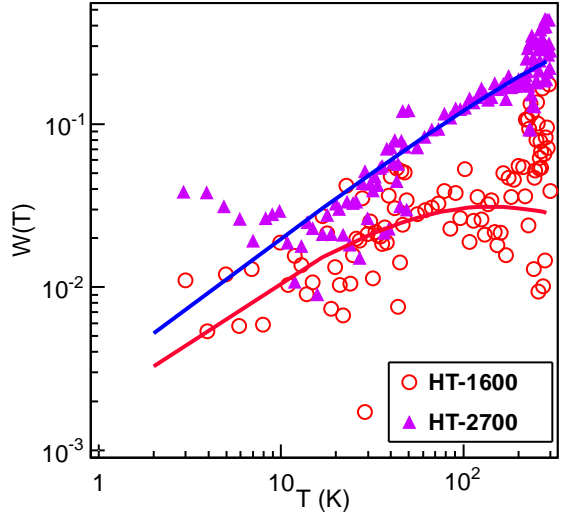}
\caption{\label{fig:W(T)}The reduced activation energy $W(T)$
versus temperature $T$ for nanocarbon powders heat
treated at $1600$ and $2700^{\circ}\mathrm{C}$. The solid lines
are analytical functions calculated for the reduced activation energy
$W(T)$. }
\end{figure}

The reduced activation energy $W(T)$ may reveal important
information about the nature of the transport in the sample, and it is
defined as \cite{Vora,Vavro}

\begin{equation}
W(T)=-\frac{d\ln\left(\rho\left(T\right)\right)}{d\ln\left(T\right)},\label{eq:activation}
\end{equation}
where $\rho(T)$ is the resistivity dependence on temperature $T$.
We have determined the reduced activation energies numerically from
the experimental data in figure \ref{fig:RvsT} and also analytically
from equation \ref{eq:res_1600}.
The analytical results are: (i)  $W(T)=\beta R_{0}z/\left\{ \left[R_{r}+R_{0}/(1+z)\right]\left(1+z\right)\right\} $
for the HT-1600 sample, and (ii) $W(T)=\beta z/\left(1+z\right)$
for the HT-2700 sample, where $z=\left(T/T_{0}\right)^{\beta}$.
The results of the numerical data analysis (symbols)
and analytical expressions (solid lines) are shown in figure \ref{fig:W(T)}.

\subsection{\label{sub:Conductance-G-versus-T}Measurements of conductance versus
magnetic field}

The sample conductance $G(B)$ versus magnetic fields in the range
$-9\leq B\leq9\:\mathrm{T}$ was measured at different temperatures
$2\leq T\leq300\:\mathrm{K}$. We have analyzed the change of conductance
$\Delta G=G(B)-G_{0}$ and relative change of conductance $\Delta G/G_{0}$,
where the conductance $G_{0}$ is the sample conductance without magnetic
field.

We followed the approach by Vavro\textit{ et al.} \cite{Vavro} to scale
magetoresistance data using a universal scaling form $Af\left(B/B_{\phi}\right)$.
$A$ is the amplitude and $B_{\phi}$ is the magnetic field that induces
one magnetic flux quantum $\Phi_{o}$ though a weak localization (WL) electron scattering
loop. Here, $\Phi_{0}=h/2e$, with $h$ being Planck's
constant and $e$ the electron charge.
We found that a similar universal scaling equation
\begin{equation}
f(cx)=c^{\alpha}f(x),\label{eq:univ_scaling}
\end{equation}
where $f(x)$ is the relative conductance $\Delta G/G_{0}$, $c$
is a positive constant, and $\alpha$ is a scaling exponent, can be
used for our data.

\begin{figure}
\includegraphics[width=7.5cm]{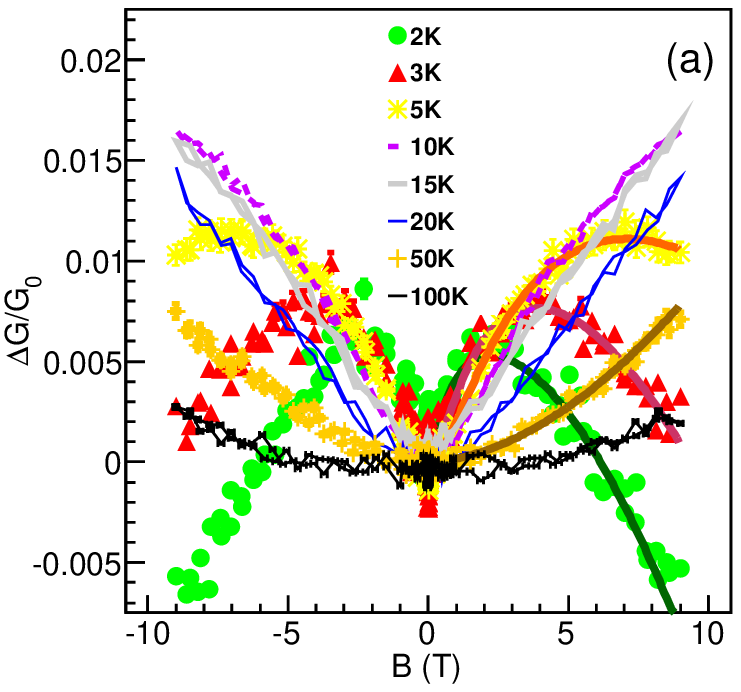}
\includegraphics[width=7.5cm]{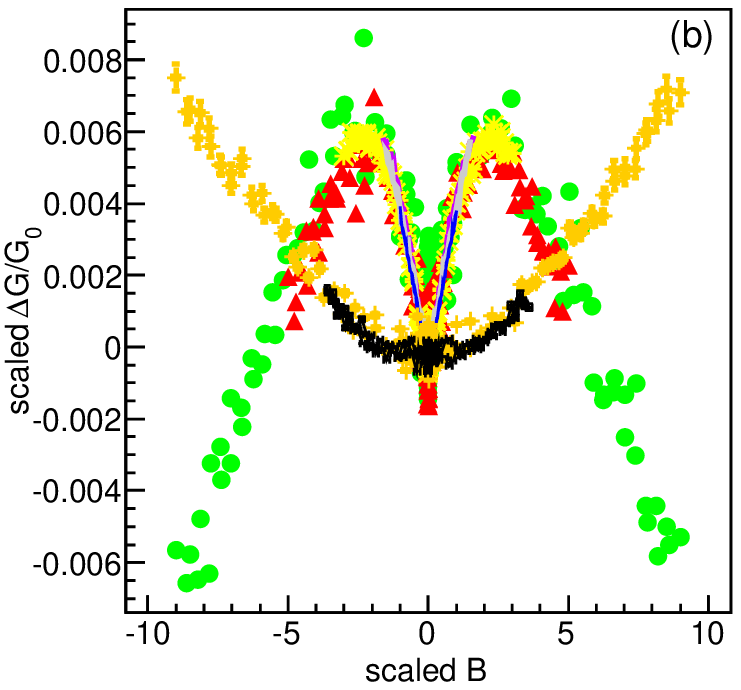}
\caption{\label{fig:GH_1600}(a) The change of conductance $\triangle G$
versus magnetic field $B$ of carbon powder HT at $1600^{\circ}\mathrm{C}$.
The graphs for temperature $T=2,\:3$ and $5\:\mathrm{K}$ are approximated
by equation (\ref{eq:del_sigma}) with the fitting parameters presented in
table \ref{tab:B_1600}. The data for $T=50\:\mathrm{K}$ (dark yellow)
are approximated by the equation $\Delta G\propto B{}^{\delta}$ with
$\delta=1.80\pm0.01$, shown as a brown solid line. (b) Using equation
(\ref{eq:univ_scaling}) the experimental data scale into one single
curve for low temperatures $2\leq T\leq20$. For temperatures $50\leq T\leq100\:\mathrm{K}$
the same scaling collapsed the data into a second, different curve. The temperature symbols
have the same meaning as in (a).}
\end{figure}

The magnetoconductance of the HT-1600 sample in figure \ref{fig:GH_1600}
(a) is positive, $\Delta G>0$, at low temperatures $2\leq T\leq5\:\mathrm{K}$
and low magnetic fields $|B|<6\:\mathrm{T}$. The increase of sample
conductance with increasing applied magnetic field is a signature
of WL \cite{Lee1985}. We found that
the graphs of $\Delta G/G_{0}$ versus $B$ for the HT-1600 sample
scale into two different curves as displayed in figure \ref{fig:GH_1600}
(b). Here, the universal scaling equation (\ref{eq:univ_scaling})
with the best-fit exponent $\alpha=0.6$ was applied. For temperatures
$2\leq T\leq20\:\mathrm{K}$ one scaling curve was found, and for
$50\leq T\leq100\:\mathrm{K}$ a second curve was obtained.

We have found that the relative change of conductance $\Delta G/G_{0}$
in figure \ref{fig:GH_1600} (a) can be approximated by a power law
$\Delta G/G_{0}\propto B^{\delta}$, where $0.5<\delta<1$ for temperatures
$2\leq T<20\:\mathrm{K}$, $\delta\doteq1$ for $T=20\:\mathrm{K}$,
and $\delta>1$ for $20<T\leq100\:\mathrm{K}$. For example, in figure
\ref{fig:GH_1600} (a) for temperature $T=50\:\mathrm{K}$ the exponent
$\delta=1.80\pm0.01$. These values of $\delta$ are different from
the expected value $\delta=2$ (see section \ref{sec:Discussion}).

\begin{figure}
\includegraphics[width=7.5cm]{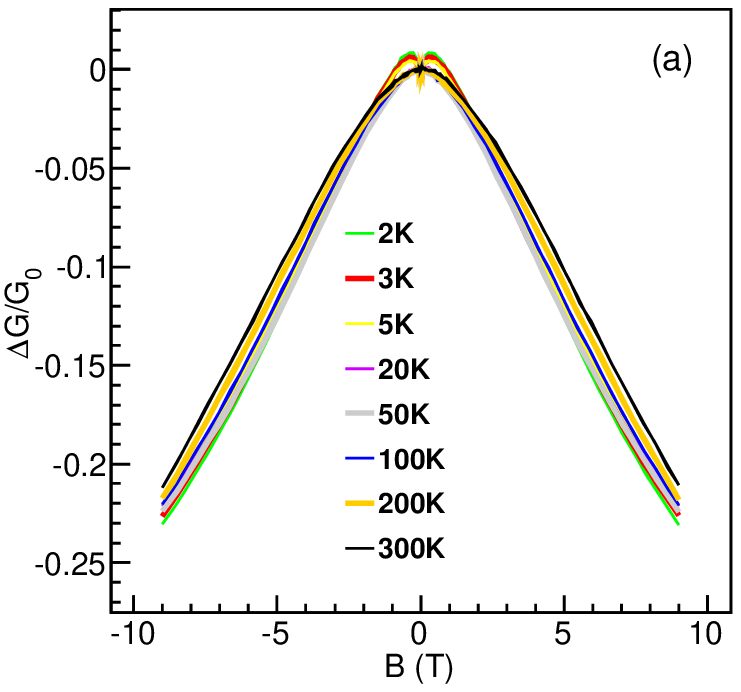}
\includegraphics[width=7.5cm]{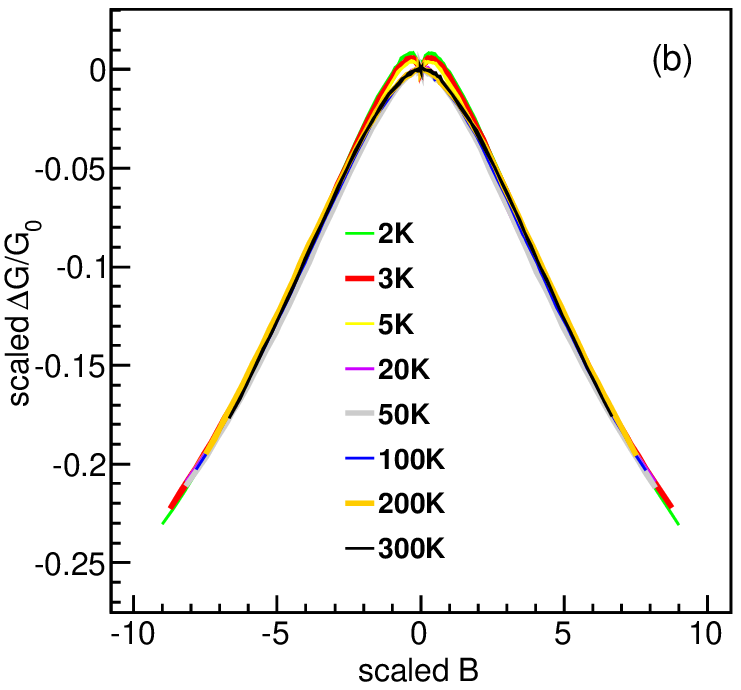}
\caption{\label{fig:GH_2700}(a) Relative change of conductance
$\triangle G/G_{0}$ versus magnetic field $B$ for the HT-2700 sample.
(b) The experimental data scale into a single curve except data measured
at low temperatures $2\leq T\leq5\:\mathrm{K}$ and low fields $\left|B\right|<3\mathrm{T}$. }
\end{figure}

\begin{figure}
\includegraphics[width=7.56cm]{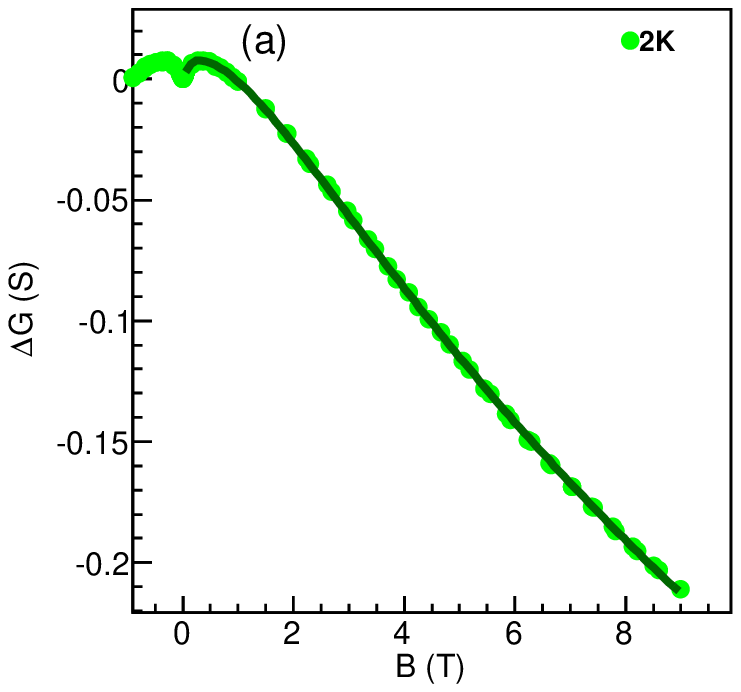}
\includegraphics[width=7.56cm]{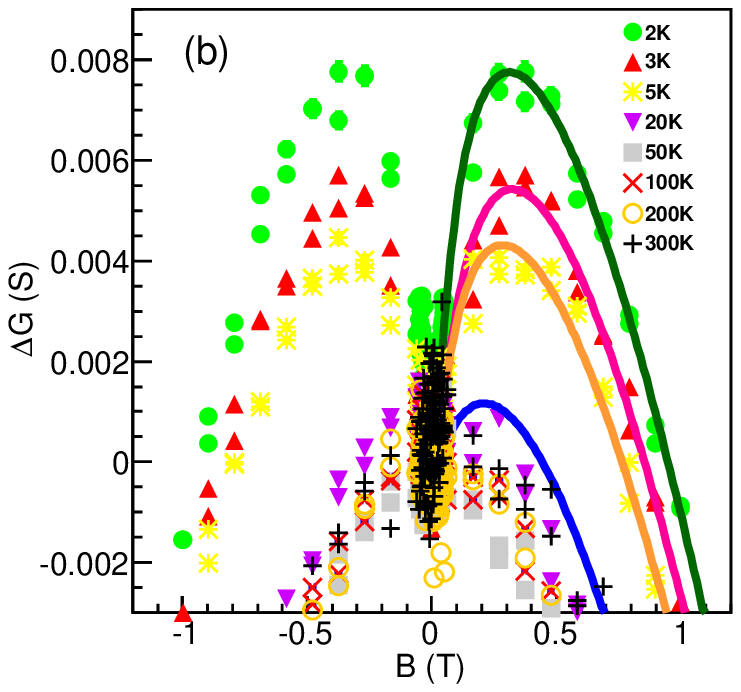}
\caption{\label{fig:WL_2700}(a) The change of conductance $\Delta G$ versus
magnetic field $B$ at $T=2\:\mathrm{K}$ where the experimental data
are approximated by equation (\ref{eq:del_sigma}) (dark green, solid line).
The fitting parameters are shown in table \ref{tab:B_2700}. (b) Weak
localization and antilocalization at low temperatures $2\leq T<20\:\mathrm{K}$
and antilocalization at temperatures $T>20\:\mathrm{K}$ at low fields
for the sample containing carbon powder HT-2700. The graphs for temperatures
$T=2,\:3,\:5$ and $20\:\mathrm{K}$ are approximated by equation (\ref{eq:del_sigma}).
The parameters are shown in table \ref{tab:B_2700}.}
\end{figure}

In a similar way, the relative change of conductance $\Delta G/G_{0}$ for the HT-2700 sample,
presented in figure \ref{fig:GH_2700} (a) could be rescaled
in the broad temperature range $2\leq T\leq300\:\mathrm{K}$.
In this case, as seen in figure \ref{fig:RvsT} (b),
the effect of the applied magnetic field was much stronger and the maximal relative change of
conductance in was about $\Delta G/G_{0}=0.23$. Quite good data collapse using scaling
equation (\ref{eq:univ_scaling}) was found giving the same exponent as before,
$\alpha=0.6$. Here, almost all data scale into the same single curve,
figure \ref{fig:GH_2700} (b), except data at low temperatures $2\leq T\leq5\:\mathrm{K}$
and low magnetic fields $\left|B\right|<3\:\mathrm{T}$.
The magnetoconductance  of this sample shows $\Delta G/G_{0}>0$
for temperatures $2\leq T\leq20\:\mathrm{K}$ and magnetic fields
$\left|B\right|<0.9\:\mathrm{T}$  (figure \ref{fig:WL_2700} (b)) which is interpreted as WL.

Surprisingly, we have found that the equations developed for the WL
in graphene \cite{McCann} can approximate the data shown in
figure \ref{fig:GH_1600} and \ref{fig:GH_2700}. We have fitted
our data to the following expression for the relative change of magnetoconductance

\begin{equation}
\Delta G=d\Biggl[F\left(\frac{B}{B_{\phi}}\right)-F\left(\frac{B}{B_{\phi}+2B_{i}}\right)-2F\left(\frac{B}{B_{\phi}+B_{i}+B_{*}}\right)\Biggr]
\label{eq:del_sigma}
\end{equation}
where $F\left(z\right)=\ln\left(z\right)+\psi\left(1/z+1/2\right)$,
$\psi\left(x\right)$ is the digamma function, $B_{\phi}$, $B_{i}$,
and $B_{*}$ are characteristic fields related to the various electron
scattering processes in the material, and $d$ is a constant. For
graphene \cite{Tikhonenko2008,McCann,Tikhonenko2009,Pal2012,Kozikov2012} the constant
$d$ is universal $d=2e^{2}/h$, with $h$ being the Planck's constant
and $e$ the electron charge.
It was found that $\tau_{\phi}^{-1}=4DeB_{\phi}/\hbar$ for the inelastic
decoherence time $\tau_{\phi}$, where $D$ is the diffusion coefficient,
and $\hbar=h/2\pi$. Similar expressions were valid for the intervalley
scattering time $\tau_{i}$ and a combined scattering time $\tau_{*}$
\cite{McCann}. These characteristic times can also be related to characteristic
electron scattering lengths given by
\begin{equation}
L_{\phi}=\sqrt{D\tau_{\phi}}=\sqrt{\frac{\hbar}{4eB_{\phi}}},\label{eq:Lphi}
\end{equation}
and similar expressions for the other length scales $L_{i}$ and $L_{*}$ with $B_{i}$ and $B_{*}$, 
respectively, replacing $B_{\phi}$ in equation (\ref{eq:Lphi})
\cite{McCann,Cai}.
In our case, $d$ in equation (\ref{eq:del_sigma}) is a parameter to
approximate the experimental data and is related to the effective
number of electron transmission channels \cite{Duan-Guojun}.

The equation (\ref{eq:del_sigma}) was used to approximate and parametrize
the experimental data for the HT-1600 sample in figure \ref{fig:GH_1600}
(a) and data for the HT-2700 sample shown in figure \ref{fig:WL_2700}.
For low temperatures $T=2,\:3$ and $5\:\mathrm{K}$ we have
determined the characteristic $B$-fields, which are presented in
tables \ref{tab:B_1600} and \ref{tab:B_2700} for the HT-1600 sample and HT-2700 sample, respectively.
For HT-2700 data where fitted only for low magnetic fields $|B|<1\:\mathrm{T}$
except for the $T=2K$ data shown in figure \ref{fig:WL_2700} (a) which showed
excellent agreement over the whole range of magnetic fields $0.01<B<9\mathrm{\: T}$.
The values of the parameter $d$ from equation (\ref{eq:del_sigma})
were determined as $d_{1600}=0.882\pm0.032\;\mathrm{S}$ and
$d_{2700}=0.1558\pm0.0008\:\mathrm{S}$. The resulting
fits are shown as solid lines superimposed on the 2, 3 and 5 K data
sets in figures \ref{fig:GH_1600} (a), \ref{fig:WL_2700} (a) and \ref{fig:WL_2700} (b).

\begin{table}
\caption{\label{tab:B_1600}The characteristic field parameters for the HT-1600
sample at low temperatures calculated from fits to equation (\ref{eq:del_sigma}).}
\begin{tabular}{cccc}
\hline
\noalign{\vskip\doublerulesep}
T (K) & $B_{\phi}$ (T) & $B_{i}$ (T) & $B_{\star}$ (T)\tabularnewline[\doublerulesep]
\hline
2 & $1.70\times10^{-1}$ & $6.46\times10^{-4}$ & $21.5$\tabularnewline
3 & $4.12\times10^{-1}$ & $2.67\times10^{-3}$ & $25.8$\tabularnewline
5 & $7.83\times10^{-1}$ & $7.40\times10^{-3}$ & $40.7$\tabularnewline
\hline
\end{tabular}
\end{table}

\begin{figure}
\includegraphics[width=7.5cm]{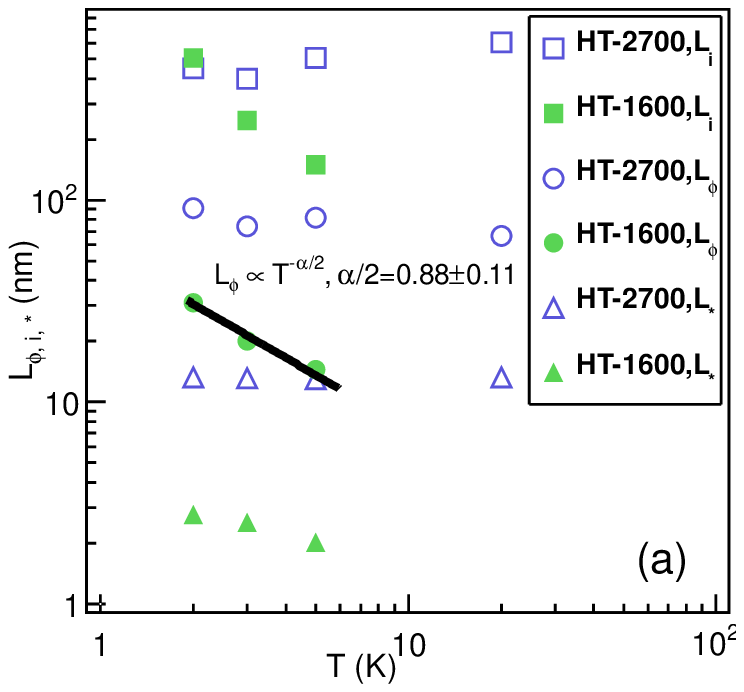}
\includegraphics[width=7.5cm]{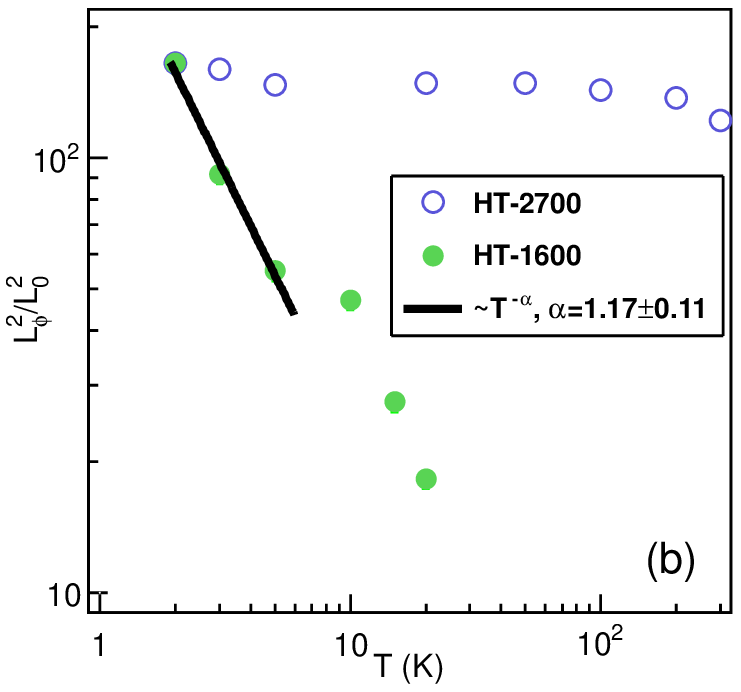}
\caption{\label{fig:LT} (a) Characteristic scattering lengths
$L_{\phi,i,\ast}$ versus temperature $T$ for HT-1600 and HT-2700
nanocarbon powders. These lengths were determined using the fit parameters
of the conductance versus magnetic field analysis of tables \ref{tab:B_1600}
and \ref{tab:B_2700}. For the HT-1600 sample the dephasing length
is approximated by a power law $L_{\phi}\propto T^{-\alpha/2}$, where
the exponent $\alpha=1.76\pm0.22$. (b) The square of the normalized, inelastic
scattering length $L_{\phi}^{2}/L_{0}^{2}$ versus temperature $T$
for the HT-1600 and HT-2700 nanocarbon powders as extracted
from the scaling function equation (\ref{eq:univ_scaling}). For the HT-1600
sample, the fit to a power law $L_{\phi}^{2}\propto T^{-\alpha}$, where
$\alpha=1.17\pm0.11$, is shown as a solid line. In this method, the
$L_{0}$ parameter is undetermined and sample specific.}
\end{figure}

According to the results of table
\ref{tab:B_1600} and equation (\ref{eq:Lphi}), the characteristic scattering
lengths for HT-1600 are in the ranges $14<L_{\phi}<31\:\mathrm{nm}$, $148<L_{i}<504\:\mathrm{nm}$,
and $L_{*}=2.4\pm0.4\:\mathrm{nm}$.
Similarly, for the HT-2700 sample the values calculated from table \ref{tab:B_2700}
are in the ranges $66<L_{\phi}<92\:\mathrm{nm}$,
$398<L_{i}<603\:\mathrm{nm}$, and $L_{\ast}=13.05\pm0.15\:\mathrm{nm}$
These values of $L_{\phi,i,*}$ are presented in figure \ref{fig:LT} (a).

\begin{table}
\caption{\label{tab:B_2700}The characteristic fields for HT-2700 sample at
low temperatures calculated from fits to equation (\ref{eq:del_sigma}).}
\begin{tabular}{cccc}
\hline
\noalign{\vskip\doublerulesep}
T (K) & $B_{\phi}$ (T) & $B_{i}$ (T) & $B_{\star}$ (T)\tabularnewline[\doublerulesep]
\hline
2 & $1.98\times10^{-2}$ & $8.17\times10^{-4}$ & $0.94$\tabularnewline
3 & $2.97\times10^{-2}$ & $1.03\times10^{-3}$ & $0.95$\tabularnewline
5 & $2.46\times10^{-2}$ & $6.45\times10^{-4}$ & $0.99$\tabularnewline
20 & $3.71\times10^{-2}$ & $4.53\times10^{-4}$ & $0.95$\tabularnewline
\hline
\end{tabular}
\end{table}

From the scaling equation (\ref{eq:univ_scaling})
we can determine the parameter $c=1/B_{\phi}$ that collapses the experimental
data in figures  \ref{fig:GH_1600}(b) and \ref{fig:GH_2700}(b).
According to equation (\ref{eq:Lphi}) $B_{\phi}\propto L_{\phi}^{-2}$,
and we then find $L_{\phi}^{2}$ from the relation $L_{\phi}^{2}/L_{0}^{2}=B_{0}/B_{\phi}$,
where $L_{0}$ and $B_{0}$ are undetermined normalization constants.
As before, $B_{\phi}$ is the field related to the effect of one magnetic
flux quantum $\Phi_{0}$ and $L_{\phi}$ is the corresponding phase
coherence length. The resultant values for the square of the normalized phase coherent
length $L_{\phi}^{2}/L_{0}^{2}$ versus $T$ for both samples based on the data scaling
are shown in figure \ref{fig:LT} (b).

Whereas the phase coherence length $L_{\phi}$ of the
HT-2700 sample shows only a weak temperature dependence, for the HT-1600
sample the temperature dependence follows $L_{\phi}\propto T^{-\alpha/2}$
with the exponent extracted from the fitted conductance parameters, $\alpha=1.76\pm0.22$, being
somewhat higher than that found from the scaling analysis, $\alpha=1.17\pm0.11$.
These fits for the exponents are shown as solid lines in figures \ref{fig:LT} (a) and \ref{fig:LT} (b).

\section{\label{sec:Discussion}Discussion}

The investigated composite samples are disordered macroscopic objects consisting
of disk- and cone- shaped nanocarbon particles which are randomly oriented in sample.
The particle volume fraction in the samples  (section \ref{sub:Polymerized-carbon-powder})
was high and well above the percolation threshold \cite{Stankovich} to
ensure many conducting paths through the macroscopic sample. The transport properties of these samples
are reproducible.

TEM images of single grains or particles show that the crystalline
quality is considerably improved on heat treatment. It should be noted
that the TEM images in figure \ref{fig:TEM} only shows the layering
structure near the rim of the particles where the layers bend and
are perpendicular to the disk surface \cite{Hage-1}. The nanocrystalline
structure may then be much better away from the edges and near the
core of the particles \cite{Garberg}.

The resistivities of samples  at temperature $T=2\:\mathrm{K}$,
$\rho_{1600} = 6.9\times10^{-4}\:\Omega\mathrm{m}$ and $\rho_{2700} = 5.45\times10^{-4}\:\Omega\mathrm{m}$,
are higher than the typical critical resistivity of
metals $\rho\approx1\times10^{-6}\:\Omega\mathrm{m}$ \cite{Imry} and
are also higher than Mott's criterion \cite{Imry,Vavro}
for minimum metallic conductivity, which corresponds to
$\rho\left(\mathrm{metal}\right)<\approx5\times10^{-5}\:\Omega\mathrm{m}$.
In such cases a direct application of quasiclassical theories is not
possible and quantum corrections are needed \cite{Imry}.

As shown in figure \ref{fig:W(T)}, for the temperature range $10\leq T\leq300\:\mathrm{K}$,
the reduced activation energy $W(T)$ increases with $T$, which shows that  both
samples belong in the metallic regime \cite{Vavro}. For the HT-2700 sample there
is a trend of $W(T)$ decreasing with $T$ below about 7 K, which may be interpreted as
crossover toward an insulating type of behaviour. The wide scattering
of the points for the HT-1600 sample prohibits any conclusion about
possible shift in the behaviour at low temperature for this sample.

At low temperatures,  $2\leq T\leq40\:\mathrm{K}$, the changes in resistance with temperature are
relatively small, as quantified by the paramater
$\gamma=\rho\left(1.9\:\mathrm{K}\right)/\rho\left(40\:\mathrm{K}\right)$.
We find $\gamma_{1600}=1.04$, $\gamma_{2700}=1.08$.
If  $\gamma<2$ the sample belongs in a metallic regime
and the density of carriers is high \cite{Sarma}. The data could not be fitted
to any exponential temperature dependence. This implies that our experimental data cannot be discussed
within the frame of variable-range hopping transport \cite{Imry,Vavro}. Similar properties
were observed in a single graphite microdisk \cite{Dujardin2001} and for nanocrystalline graphite \cite{Mandal2013}.

In zero magnetic field, the temperature dependence of the resistance, $R(T)$, 
of the HT-1600 sample shown in figure \ref{fig:RvsT}(a) and of the HT-2700
sample in figure \ref{fig:RvsT}(b) are approximated by the same equation
(\ref{eq:res_1600}). In these samples inelastic scattering events
due to interfering paths of electrons may give a contribution to the
resistance at finite $T$, which is absent for finite magnetic
fields $H>0$ \cite{Lee1985,Vavro}.

These resistance behaviours reveal 3D electron transport
for temperature $30<T<300\:\mathrm{K}$. We initially considered that
Kawabata's theory of negative MR \cite{Kawabata} in 3D might explain
the results shown in figures \ref{fig:GH_1600} and \ref{fig:GH_2700}.
The theory predicts asymptotic forms $\Delta G\propto\sqrt{B}$ for
large magnetic fields and $\Delta G\propto B^{2}$ for small magnetic
fields \cite{Lee1985,Vora,Vavro,Kawabata}. The experimental data for
the two samples shown in figures \ref{fig:GH_1600} and \ref{fig:GH_2700}
could not be approximated by the the asymptotic forms of the Kawabata
equation neither in weak nor in strong magnetic fields (see section \ref{sub:Conductance-G-versus-T}).
We have also verified that these data
cannot be modeled using the theory of Hikami \textit{at al} \cite{Hikami}.

The observed layered structure of the nanocarbon particle
and the earlier finding that 3D carbon conductors can behave as
2D systems \cite{Vavro} but different from WL in 2D electron gas \cite{Altshuler1980,Hikami},
inspired us to look into models used for electron transport in 2D
multilayer graphene \cite{SevakSingh2012,Friedman2010}. We have successfully
applied the equation (\ref{eq:del_sigma}), which was originally derived
to explain transport in single layer graphene \cite{McCann}, to approximate
the results shown in figures \ref{fig:GH_1600} and \ref{fig:GH_2700}.
Magnetoconductivity in figure  \ref{fig:GH_2700}
resembles the results presented for multilayer epitaxial graphene
reported by Friedman \emph{et al.} \cite{Friedman2010} and by Singh
\emph{et al.} \cite{SevakSingh2012}. Since our samples at $T=2\,\mathrm{K}$
have conductivities that are several tenfold times the quantum conductivity
$\sigma_{0}=2e^{2}/h$, they consist of very many conducting paths
that each may be partly graphene-like \cite{Baker2012,SevakSingh2012,Wu2007,Friedman2010,Lara-Avila2011}.
For example, at similar $B-T$ conditions the samples reported by Singh \emph{et al.} \cite{SevakSingh2012}
consisting of about $20$ graphene layers show thousand times lower
conductivity then our HT-2700 sample in figure \ref{fig:WL_2700}.
After conductance normalization (equation \ref{eq:del_sigma}) the experimental data are well approximated
by the McCann model \cite{McCann} for a wide range of temperatures
and magnetic fields. Our magnetoconductivity results in figure \ref{fig:WL_2700}
(a) may be compared to experimental data for graphene \cite{Tikhonenko2008},
rippled graphene \cite{Lundeberg2010}, bilayer graphene \cite{Gorbachev2007}
or multilayered graphene \cite{SevakSingh2012}. It may be noted that
the model for bilayer graphene \cite{Gorbachev2007} which has a positive
sign of the third term in equation (\ref{eq:del_sigma}), clearly did not
fit our experimental data.

Typical sizes of most of the nanocarbon particles
were $\sim0.3-3.0\,\mu\mathrm{m}$ with the thickness of the walls
in the range $10-50$ nm. This thickness of our  nanoparticles is comparable
but slightly larger than the thickness of multilayered epitaxial graphene
samples \cite{SevakSingh2012,Friedman2010}. However, although their
remaining external dimensions are essentially smaller than the dimensions
of the multilayer samples \cite{SevakSingh2012,Friedman2010}, the
grain size of our material may be approaching that of multilayer
graphene. The calculated dephasing lengths $L_{\phi}$ (figure \ref{fig:LT} (a))
are smaller than the in-plane carbon particle sizes but comparable or larger
than typical particle thickness. For example, the typical lengths
$L_{\phi}$ in HT-1600 ($14<L_{\phi}<31\:\mathrm{nm}$) and HT-2700
($66<L_{\phi}<92\:\mathrm{nm}$) are smaller than the diameter of
the disk- or cone-shaped particles. These are reasons why our nanosized
objects, graphite-like nanosize  crystalities \cite{Romanenko2006}, 
and the multilayer graphene samples \cite{SevakSingh2012,Friedman2010}
can be considered to be quasi-2D objects. The intervalley $L_{i}$
and intravalley $L_{\ast}$ scattering lengths found here are comparable
to those found in graphene \cite{Tikhonenko2008,Pal2012}. The dephasing
length $L_{\phi}$ is smaller than the values reported for graphene \cite{Tikhonenko2008,Pal2012}
and multilayer graphene \cite{SevakSingh2012}. We have found that
$L_{\phi}<L_{i}$, which is not typical for graphene.

The temperature dependence of dephasing length $L_{\phi}$
can be determined using an alternative approach based on magnetoconductivity
scaling \cite{Vora,Vavro}. For all samples  the magnetoconductivities
at different temperatures (figures  \ref{fig:GH_1600}(b)
and \ref{fig:GH_2700}(b)) scale using the universal scaling form equation
(\ref{eq:univ_scaling}) with a single parameter $c=1/B_{\phi}$. We
found that our samples follow a scaling relation \cite{Vora,Vavro}
$\Delta G/G_{0}=-Af\left(B/B_{\phi}\right)$ with $A=\left(1/B_{\phi}\right)^{0.6}$,
which give rise to the excellent data collapse in figure \ref{fig:GH_1600}(b).
In contrast to the previous results \cite{Vora,Vavro}, our data do
not scale onto a single universal curve. For the HT-1600 sample
we found two scaling functions which divide the investigated temperature
range into low ($T<T_{C}$) and high ($T>T_{C}$) temperature intervals.
The critical temperature falls in the range $30<T_{C}<50\:\mathrm{K}$.
For the HT-2700 sample we found that except data at low temperatures
$T<20\:\mathrm{K}$ and low magnetic fields $B<3\:\mathrm{T}$ all
remaining data fall onto one universal function. The deviations from
universal scaling forms are a consequence of WL effects.

From the scaling form we determined $B_{\phi}$, or equivalently,
the square of the phase coherence length $L_{\phi}$ since $B_{\phi}^{-1}\propto L_{\phi}^{2}$.
The temperature dependence of $L_{\phi}^{2}$ reveals information
about electron scattering mechanisms. The $B_{\phi}$-field vs. temperature
$T$ at low temperatures obeys the scaling $B_{\phi}^{-1}\propto L_{\phi}^{2}\propto\tau_{\phi}\propto T^{-\alpha}$.
In figure \ref{fig:LT} (b) we determined the scaling exponent $\alpha=1.17\pm0.11$
for  the  HT-1600 samples. For HT-1600 sample and temperature
$2\leq T\leq5\:\mathrm{K}$ (figure \ref{fig:LT} (a)) a slightly
higher value, $\alpha\approx1.76$, was determined based on the fits
to equation (\ref{eq:del_sigma}). Considering these two methods to be equivalent,
we find the scaling exponent of the HT-1600 sample to be in the range
$\alpha=1.2-1.8$. The scaling exponent of  the HT-1600 sample is then near the prediction of the
exponent $\alpha=1.5$ for \textit{e-e} scattering in the dirty limit \cite{Vora,Menon}.
However, experiments on twisted bilayer graphene \cite{Meng2012} found
that the dephasing rate was dominated by electron-electron Coulomb
interaction with $\tau_{\phi}^{-1}\propto T^{2}$, i.e. $\alpha=2$.
For the HT-2700 sample, the coherence lengths $L_{\phi}$ determined
using the two different methods show only a weak temperature
dependence, as seen in figure \ref{fig:LT}. 
Similar results were observed in a graphite microdisk \cite{Dujardin2001}
and in multilayer epitaxial graphene on SiC \cite{SevakSingh2012,Friedman2010}.
This  property  was attributed to electronically decoupling of graphene layers
\cite{SevakSingh2012,Friedman2010}.

The band structure around the Dirac points in monolayer graphene is
linear while in bilayer AB stacked graphene it is quadratic \cite{Castro,Gorbachev2007}.
Multilayer regularly ABA or ABC stacked graphene show a
band structure which depends on the number of graphene layers, but
for number of layers $n>11$ the band structure is similar to graphite \cite{Partoens2006}.
However, linear band spectrum is preserved when graphene layers are
misoriented \cite{LopesdosSantos2007,Latil2007,Ohta2007,Hass2008},
i.e., adjacent rotated planes become electronically decoupled \cite{Hass2008}.
In our nanoparticles faceting angles of $\sim22^{\circ}$ were observed \cite{Garberg},
which fits well with the second commensurate rotation $\theta_{2}=21.7^{\circ}$
found in the calculations by Loopes dos Santos \textit{et al.} \cite{LopesdosSantos2007}.
Electronically decoupled graphene layers were observed in multilayer
graphene \cite{SevakSingh2012,Friedman2010} and recently very high mobilities
in multilayer 2D samples that look similar to ours were reported
and interpreted to be a consequence of electronic decoupling of turbostratic graphene
layers \cite{Hernandez2013}. In those samples the reported mobilities of inner layers
may approach that of suspended graphene. These results support the idea that nearly
non-interacting parallel graphene layers may exist in several types of multilayered graphene.
Thus, the model of decoupled graphene layers used to approximate magnetoconductivities
in equation (\ref{eq:del_sigma}) may be justified in the present case.

Electron interactions in certain multilayer
graphene samples, e.g., our HT-2700 sample in figures \ref{fig:LT} 
and the epitaxial graphene of Singh et al. \cite{SevakSingh2012},
do not behave conventional because the dephasing length $L_{\phi}$
is almost independent of temperature. In graphene flakes the electron
dephasing rate obeys the usual linear $T$-dependence $\tau_{\phi}^{-1}\propto T$ \cite{Tikhonenko2008}.
Tikhonenko \emph{et al.} \cite{Tikhonenko2008} concluded that electron
interference in graphene is significantly different from other 2D
systems, however \textit{e-e} interaction does not show unconventional
behaviour. Analyzing our results and the results of other groups one
may conclude that unconventional temperature behaviour of dephasing
length $L_{\phi}$, or dephasing time $\tau_{\phi}$, is a typical
property of turbostratic-like graphene layers. To explain this specific
behaviour a new theoretical approaches will be needed. Very recent
models of twisted bilayer graphene \cite{Sarrazin} have predicted
novel effects such as exciton swapping between sheets.

In the case of inelastic electron scattering \cite{Lee1985} magnetoconductivity
measurements should be consistent with resistivity versus temperature
measurements. The exponent $\beta$ in equation (\ref{eq:res_1600}) depends
on scattering mechanism. For $L_{\phi}^{2}\propto T^{-\alpha}$, it
is expected that $\beta=\alpha/2$. Here, the $\beta$ values of our
samples are found in table \ref{tab:Rfit} and $\alpha$ values
are shown in figures \ref{fig:LT} (a) and \ref{fig:LT} (b). The HT-1600
sample shows the best consistency between the these exponents, $\alpha\approx1.76$
and $\beta\approx0.84$. The exponents of the  HT-2700 samples
are not consistent with inelastic electron scattering in 3D \cite{Lee1985}.

\section{\label{sec:Conclusions}Conclusions}

Electronic transport in macroscopic bulk composite samples containing nanocarbon 
disks and cones in PMMA has been investigated using magnetotransport measurements
in order to find characteristic electron scattering lengths and their temperature 
dependencies. We found that the magnetotransport properties are strongly dependent 
on the increase in the fraction of crystalline phases in the nanocarbon particles
after heat treatment of the particles. We applied the McCann 
theory of magnetoconductivity of single layer graphene to approximate the 
low temperature data and interpret the changes of magnetoconductivity for 
the nanocarbon heat-treated at $1600^{\circ}$ C as a result of temperature variations
of the electron scattering length $L_{\phi}$. We found the exponent $\alpha$ in the 
range 1.17-1.76 for the temperature dependence of $L_{\phi}$,  $L_{\phi}^2\propto T^{-\alpha}$. 
The material heat-treated at $2700^{\circ}$ C did not show any such clear changes at low temperatures.
The characteristic electron scattering lengths are typically less than about 100 nm, 
much smaller than the particle sizes.

\ack{The authors would like to thank J. Voltr of Czech Technical University
in Prague for measuring the Fe contamination of the samples, J.  P.
Pinheiro of n-Tec AS for providing the samples used in this study, and Y.
Galperin and J.  Bergli of University of Oslo for useful discussions.  This
work was supported by SAS Centre of Excellence: CFNT MVEP and by Research
Council of Norway grant no.  191621/F20,
University Science Park TECHNICOM for Innovation Applications Supported
by Knowledge Technology, ITMS: 26220220182,
supported by the Research \& Development Operational Programme funded by the ERDF}.

\end{document}